\documentclass[prd,twocolumn,showpacs,floatfix,amsmath,nofootinbib,amssymb,floatfix]{revtex4}
\usepackage{graphicx,color,dcolumn,booktabs,bm,multirow}
\usepackage{longtable,lscape}
\usepackage{txfonts}
\usepackage{overpic}
\usepackage{amssymb}
\usepackage{indentfirst}
\usepackage{feynmf}   
\usepackage{slashed}  
\usepackage{cases}
\usepackage{color}
\usepackage{multirow}
\usepackage{epstopdf}
\usepackage{graphicx,color,dcolumn,booktabs,bm}
\usepackage{epstopdf}
\usepackage{ulem}
\usepackage{amsmath,bm}

\usepackage{braket}

\usepackage[colorlinks, citecolor=green,anchorcolor=red,menucolor=red, linkcolor=blue,filecolor=red,runcolor=red,urlcolor=blue,frenchlinks=red]{hyperref}

\def\D{\bar D}

\begin{document}

\title{A study of the decays of  $S-$wave $\bar D^\ast K^\ast$ hadronic molecules: the scalar $X_0(2900)$ and its spin partners $X_{J(J=1,2)}$}
\author{Cheng-Jian Xiao$^1$}\email{xiaocj@ihep.ac.cn}
\author{Dian-Yong Chen$^2$}\email{chendy@seu.edu.cn}
\author{Yu-Bing Dong$^{3,4,5}$}\email{dongyb@ihep.ac.cn}
\author{Guang-Wei Meng$^1$}
\affiliation{$^1$Institute of Applied Physics and Computational Mathematics, Beijing 100088, People's Republic of China\\
$^2$ School of Physics, Southeast University, Nanjing 210094, People's Republic of China\\
$^3$Institute of High Energy Physics, Chinese Academy of Sciences, Beijing 100049, People's Republic of China\\
$^4$Theoretical Physics Center for Science Facilities (TPCSF), CAS, Beijing 100049, People's Republic of China\\
$^5$ University of Chinese Academy of Sciences, Beijing 100049, People's Republic of China}

\date{\today}
\begin{abstract}
 In this work, we investigated the decays of  the fully open-flavor tetraquark state $X_0(2900)$  which was observed by the LHCb Collaboration very recently. Here, the $X_0(2900)$ was assigned as a $S-$wave $\bar D^\ast K^\ast$ hadronic molecule with $I=0$, and the effective lagrangian approach was applied to estimate the partial decay widths. Moreover, we also predicted the decay behaviors of the other unobserved $X_{J(J=1,2)}$, which were the spin partners of the $X_0(2900)$ in the $S-$wave $\bar D^\ast K^\ast$ picture. It was pointed out that the $X_1$ state with $I=0$ was a broad state with the width more than one hundred MeV, while another $X_2$ state with $I=0$ was a narrow state with the width approaching half of that for the $X_0(2900)$. In addition, our results also showed that the $\bar D^\ast K$ mode was expected to be the dominant decay mode for both $X_1$ and $X_2$. Searching for those unobserved $X_{J(J=1,2)}$  in the future experiments might be helpful to understand the nature of $X_0(2900)$.
\end{abstract}


\maketitle

\section{Introduction}\label{sec1}
Until now, the exotic family is no longer thin due to the great efforts from the experimental side.  Traces of their existence have be found in $B=0$ meson sector, baryon sector as well as the $B=2$ dibaryon sector, namely, the tetraquark states, pentaquark states, and hexaquark states . The $X(3872)$, $D_{s0}^\ast(2317)$, $Z_c$, $P_c$ are the typical examples of the remarkable exotic states (more information can be found in the review papers~\cite{Chen:2016qju,Hosaka:2016pey,Richard:2016eis,
Clement:2016vnl,Lebed:2016hpi,Esposito:2016noz,Ali:2017jda,Guo:2017jvc,
Olsen:2017bmm,Karliner:2017qhf,Liu:2019zoy,
Brambilla:2019esw}). Concerned to the constituent quarks, most of the exotic states contain  a pair of quark-antiquark, $c\bar c$ or $u\bar u$ for instance, which makes them hidden-flavor. Besides of the hidden-flavor structure, the exotic states can be composed of fully open-flavor quarks. The first fully open-flavor exotic state, as well as the  only one before September 2020, was observed in the 2016 named $X(5568)$\cite{D0:2016mwd}. It was observed by the D0 Collaboration and was expected to be consist of $\bar bsu\bar d$, which made it obviously exotic\cite{D0:2016mwd}. The $X(5568)$ was interesting and attracted a great attention for both experimentalists  and theorists \cite{Hosaka:2016pey,Lebed:2016hpi,
Esposito:2016noz,Guo:2017jvc,Olsen:2017bmm,Liu:2019zoy,
Brambilla:2019esw}. However, the later negative results for the $X(5568)$ from other collaborations bogged down the interests of the study of fully open-flavor states\cite{Aaij:2016iev,CMS:2016fvl}.\\

The situation dramatically changed very recently, since the LHCb Collaboration reported their first amplitude analysis of the $B^+\to D^+D^-K^+$ process\cite{Aaij:2020hon,Aaij:2020ypa} and where they have to introduce one spin-0 $X_0(2900)$ state and another spin-1 $X_1(2900)$ in their model in order to describe the data. Their obtained resonance  parameters were,
\begin{eqnarray}
X_{0}(2900):&&M=2866 \pm 7 \pm 2 \mathrm{MeV}, \label{eq:mass-2866} \\ \nonumber
&&\Gamma=57\pm 12 \pm 4 \mathrm{MeV},
\end{eqnarray}
and
\begin{eqnarray}
X_{1}(2900):&&M=2904 \pm 5\pm 1 \mathrm{MeV}, \label{eq:mass-2900}\\ \nonumber
&&\Gamma=110\pm 11 \pm 4 \mathrm{MeV}.
\end{eqnarray}
Their $P$ parities were determined to be positive for the spin$-0$ state, and negative for the spin$-1$ state based on the $D^-K^+$ decay channel. Besides, the isospin $I$ was still unknown, while there were two possible assignments $I=0$ and $I=1$. Therefore, the $I(J^P)$ quantum numbers of the $X_0(2900)$ and $X_1(2900)$ were $0/1(0^+)$ and $0/1(1^-)$, respectively.\\

It should be stressed that the $D^-K^+$ final state indicated the exotic structure of the observed $X_0(2900)$ and $X_1(2900)$, e.g., $\bar cd \bar s u$ quark flavors. Therefore, the two resonances were fully open-flavor states similar to the $X(5568)$, and unambiguously differed from the conventional hadrons. Those exotic states have been explained  as the tetraquark states. The hadronic molecules and compact tetraquarks are two types of tetraquark states. In the former case, the four quarks  form two hadrons, which are bounded via the strong interaction. In the later case, the quarks form a compact structure. For the particular $\bar cd \bar s u$ structure here,  Ref.~\cite{Molina:2010tx} calculated its anti-particle in 2010, a bound $D^\ast\bar K^\ast$  decaying to $D\bar K$ . The predicted  mass, width and quantum numbers were $2848$\,MeV, 59\,MeV and $I(J^P)=0(0^+)$, respectively. Morever, the authors of Ref.~\cite{Cheng:2020nho} also predicted a $cs\bar u\bar d$ state with the mass 2850\,MeV. Besides, the charmed partners of the $X(5568)$, whose structure were $su\bar d\bar c$, were predicted  \cite{Agaev:2016lkl,Chen:2016mqt}, however, the mass $M=2550$\,MeV did not fit the present observation.\\

Stimulated by the observation of the $X_0(2900)$ and $X_1(2900)$, many theoretical analyses of the two resonances have been carried out by employing various approaches\cite{Karliner:2020vsi,Liu:2020orv,He:2020jna,
Zhang:2020oze,
Liu:2020nil,Chen:2020aos,Lu:2020qmp,Huang:2020ptc,
He:2020btl,Wang:2020xyc,Hu:2020mxp,Xue:2020vtq,
Molina:2020hde,Burns:2020epm,Agaev:2020nrc,
Albuquerque:2020ugi,Chen:2020eyu,Mutuk:2020igv,
Burns:2020xne,Dong:2020rgs}.
 Ref.~\cite{Karliner:2020vsi} and~\cite{Zhang:2020oze,Wang:2020xyc} interpreted the $X_0(2900)$ as the compact tetraquark based on the constituent quark model and  QCD sum rules, respectively. Moreover, the $X_1(2900)$ was explained as the compact tetraquark state in Refs.~\cite{Wang:2020xyc,Chen:2020aos,Xue:2020vtq,
 Molina:2020hde,Agaev:2020nrc,Mutuk:2020igv}. Applying the chromomagnetic interactions diquark configuration model, the $J^P=0^+$ resonance  was also considered as a radial excited tetraquark, while the $J^P=1^-$ one was assigned as an orbitally excited tetraquark\cite{He:2020jna}. However, a calculation based on the extended relativized quark model disfavored the tetraquark interpretation\cite{Lu:2020qmp}.

  It should be mentioned that the hadronic molecules assignments were proposed\cite{Chen:2020aos,Hu:2020mxp,Liu:2020nil,
 He:2020btl,Huang:2020ptc,Molina:2020hde,
 Agaev:2020nrc,Mutuk:2020igv}. By considering the $J^P$ quantum numbers and mass threshold, the $X_0(2900)$ was explained as the $S-$wave $\bar D^\ast K^\ast$ hadronic molecule, while the $X_1(2900)$ was explained as the $\bar D_1 K$\cite{He:2020btl} and  $P-$wave $\bar D^\ast K^\ast$ hadronic molecules\cite{Huang:2020ptc}. There was also a negative results for the $\bar D_1 K$ molecule interpretation for the $X_1(2900)$, where the author found that the potential between $\bar D_1 K$ was too weak to form any bound state\cite{Dong:2020rgs}. To explore the nature of the $X_0(2900)$ and $X_1(2900)$, the production mechanism was also analysed\cite{Chen:2020eyu}. In additions, Ref.~\cite{Liu:2020orv} considered the triangle singularity to be the origin of $X_0(2900)$ and $X_1(2900)$.\\

Whether the $X_0(2900)$ and $X_1(2900)$ were compact tetraquarks, hadronic molecules or due to kinetic effects was unclear so far. In the present work, we followed the $S-$wave $\bar D^\ast K^\ast$ interpretation for the $X_0(2900)$ with isospin $I=0$ proposed in Ref.~\cite{Hu:2020mxp,Liu:2020nil,
 He:2020btl,Molina:2020hde} to investigate its decay behaviors via the effective lagrangian approach. In particular, in the $S-$wave $\bar D^\ast K^\ast$ hadronic molecule scenario, two spin partners of $X_0(2900)$ were predicted with $J=1$ and $J=2$\cite{He:2020btl,Hu:2020mxp,Molina:2020hde}. Here  we would refer $X_1$ and $X_2$ to the $J=1$ and $J=2$ states, respectively. One should note that the $X_1$ hereafter was not the $X_1(2900)$ in Eq.~(\ref{eq:mass-2900}), while the $X_0$ corresponding to the $X_0(2900)$ in Eq.~(\ref{eq:mass-2866}). Within the same molecule scenario, we also investigated the decay behaviors of $X_1$ and $X_2$.\\

The present paper is assigned as follows.  The effective lagrangians and decays are given in the  next section. Sec.~\ref{sec:3} shows our numerical results and discussion. The summary is presented in the last section.

\section{Effective lagrangians and decays}\label{sec:2}
The effective lagrangian approach was applied to estimate the decays of experimental observed $X_0(2900)$ in the present work, where it was considered as the $S-$wave $\bar D^\ast K^\ast$ hadronic molecules with the isospin $I({X_0})=0$. Besides, the decays of the predicted $X_{J(J=1,2)}$, being the spin partners of the $X_0(2900)$ in the $S-$wave $\bar D^\ast K^\ast$ picture, were also investigated, where two possible isospins $I(X_{J(J=1,2)})=0$ and $I(X_{J(J=1,2)})=1$ were adopted for our analyses.

We firstly constructed the effective lagrangians describing the interaction between the molecular state and its components,
\begin{widetext}
\begin{eqnarray}
{\cal{L}}_{X_0}(x)&=&g_{X_0}X_1(x)\int{}dy\Phi(y^2)
    \left[
    {D}^{\ast-\mu}(x+\omega_{K^{\ast}
    \bar D^
    \ast}y)K^{\ast+}_{\mu}(x-\omega_{\bar{D}^{\ast}K^\ast}y)
    -\bar {D}^{\ast0\mu}(x+\omega_{K^{\ast}
    \bar D}y)K^{\ast0}_{\mu}(x-\omega_{\bar{D}^{\ast}K^\ast}y)
    \right]+\text{H.C.},\label{eq:lag-x0}\\
{\cal{L}}_{X_1}(x)&=&ig_{X_1}\epsilon_{\mu\nu\alpha\beta}
    \partial^\mu X_1^\nu(x)\int{}dy\Phi(y^2)
    \left[
    {D}^{\ast-\alpha}(x+\omega_{K^{\ast}
    \bar D^\ast}y)K^{\ast+\beta}(x-\omega_{\bar{D}^{\ast}K^\ast}y)
    \pm\bar {D}^{\ast0\alpha}(x+\omega_{K^{\ast}
    \bar D}y)K^{\ast0\beta}(x-\omega_{\bar{D}^{\ast}K^\ast}y)
    \right]+\text{H.C.},\label{eq:lag-x1}\\
{\cal{L}}_{X_2}(x)&=&g_{X_2}X_2^{\mu\nu}(x)\int{}dy\Phi(y^2)
    \left[
    {D}^{\ast-}_{\mu}(x+\omega_{K^{\ast}
    \bar D^
    \ast}y)K^{\ast+}_{\nu}(x-\omega_{\bar{D}^{\ast}K^\ast}y)
    \pm\bar {D}^{\ast0}_{\mu}(x+\omega_{K^{\ast}
    \bar D}y)K^{\ast0}_{\nu}(x-\omega_{\bar{D}^{\ast}K^\ast}y)
    \right]+\text{H.C.}\label{eq:lag-x2},
\end{eqnarray}
\end{widetext}
where the $\pm$ corresponding to $X_{J(J=1,2)}$ states with $I=1$ and $I=0$, respectively. The coupling constant $g_{X_{J(J=,0,1,2)}}$ can be determined by the compositeness condition\cite{Weinberg:1962hj,Faessler:2007gv,
Faessler:2007us}. The $\omega_{AB}=m_A/(m_A+m_B)$, the correlation function $\Phi(y^2)$ carries the distribution information of the components in the hadronic molecule. Within the Fourier transformation, $\Phi(y^2)=\int d^4p/(2\pi)^4e^{-ipy}\tilde\Phi(-p^2)$. It should be mentioned that the Gaussian form $\tilde\Phi(p_E^2)=exp(-p_E^2/\Lambda^2)$ was widely used to estimate the decays of hadronic molecules    \cite{Weinberg:1962hj,Faessler:2007gv,
Faessler:2007us,Dong:2009yp,Branz:2009yt,Huang:2018wgr}. In Eqs.~(\ref{eq:lag-x0})-(\ref{eq:lag-x2}), the $\Lambda$ is the model parameter related to the size of the hadronic molecule.

\begin{figure*}[hbt!]
\begin{center}
\begin{tabular}{ccc}
\includegraphics[scale=0.5]{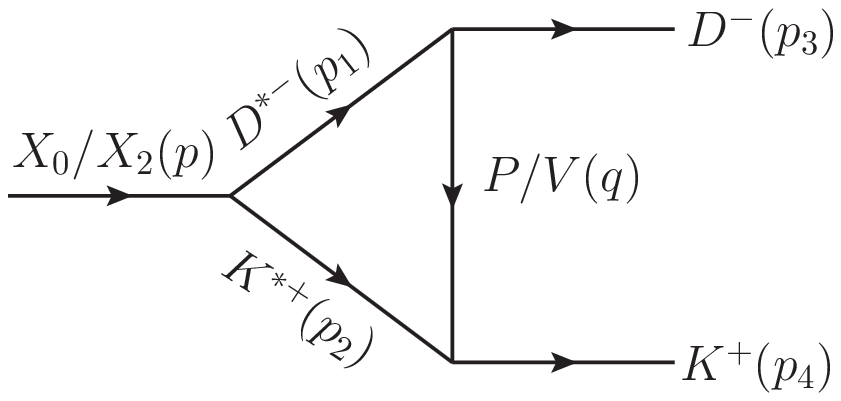}
&\includegraphics[scale=0.5]{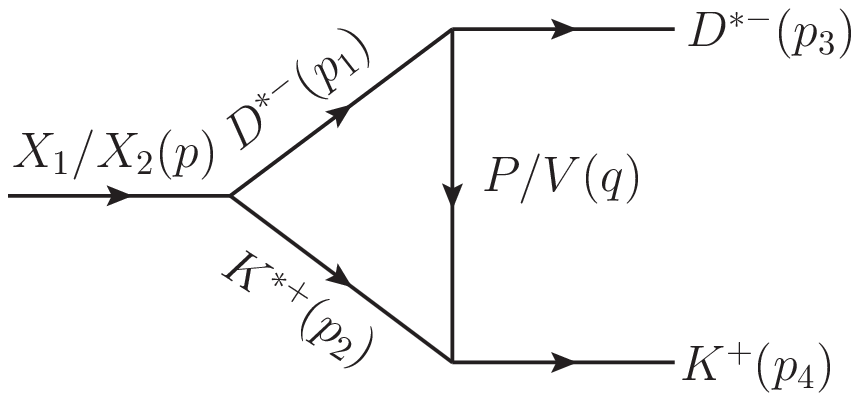}
&\includegraphics[scale=0.5]{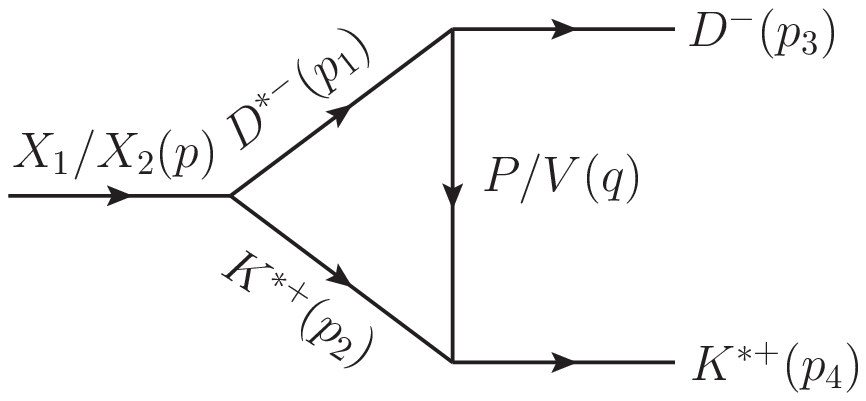}\\
(a) &(b)    &(c)
\end{tabular}
\end{center}
\caption{Diagrams contributing to the processes $X_{J(J=0,1,2)}\to D^{(\ast)-}K^{(\ast)+}$. Diagram (a) corresponding to the transition from $X_0$ to $\D^- K^-$. Diagram (b)-(c) corresponding to the transition from  $X_1$  to $\D^{\ast-} K^+$ and $\D^- K^{\ast+}$.  Diagram (a)-(c) corresponding to the transitions from $X_2$ to $\bar D^- K^+$, $\D^{\ast-} K^+ $ and $\D^- K^{\ast+}$. The $P$ and $V$ stood for the exchanged pseudoscalar mesons and vector mesons, respectively, including the $\pi^0$, $\eta$, $\eta^\prime$, $\rho^0$ and $\omega$. Besides, three additional diagrams with the intermediate $\bar D^{\ast0} K^{\ast0}$ were not presented here, which also contributed to the process $X_{J(J=0,1,2)}\to D^{(\ast)-}K^{(\ast)+}$ and were considered in the calculation.  }\label{fig:tri-x-two-body-decay}
\end{figure*}

Considering the two-body decays, the $X_0$ can decay to $\D K$, the $X_1$ can decay to $\D^\ast K$ and $\D K^\ast$, and $X_2$ can decay to $\D K$, $\D^\ast K$ and $\D K^\ast$. These transitions occured via the triangle diagrams (presented in Fig.~\ref{fig:tri-x-two-body-decay}), where the hadronic molecule and the final state are connected through the $\bar D^\ast$ and $K^\ast$ by exchanging a proper hadrons. Here, the exchanged hadrons can be either pseudoscalar meson and vector meson, including
\begin{eqnarray}
\begin{array}{cccc}
  P:    &\pi,   &\eta,  &\eta^\prime,\\
  V:    &\rho,    &\omega.
\end{array}
\end{eqnarray}

As we can see from the Fig.~\ref{fig:tri-x-two-body-decay}, the effective lagrangians describing the interaction between the charmed (strange) mesons and exchanged hadrons were essential\cite{Lin:1999ad,Oh:2000qr},
\begin{eqnarray}
&&\mathcal{L}_{D^\ast DP}=ig_{D^\ast DP}
    (D^{\ast\mu}\partial_\mu P\bar D
    -D \partial_\mu P\bar D^{\ast\mu}),\\
&&\mathcal{L}_{D^\ast D^\ast P}=
    -g_{D^\ast D^\ast P}\epsilon^{\mu\nu\alpha\beta}
    \partial_\mu D^\ast_\nu P\partial_\alpha\bar D^\ast_\beta,\\
&&\mathcal{L}_{D^\ast DV}=-g_{D^\ast DV}
    \epsilon_{\mu\nu\alpha\beta}
    D\partial^\mu V^\nu\partial^\alpha \bar D^{\ast\beta}+\text{H.C.},\\
&&\mathcal{L}_{D^\ast D^\ast V}=ig_{D^\ast D^\ast V}
    \big[D^\ast_\mu(\partial^\mu\bar D^\ast_\nu V^\nu-\partial^\mu V^\nu\bar D^\ast_\nu)\nonumber\\
&&\phantom{\mathcal{L}_{D^\ast D^\ast\rho}=}
    +(D^\ast_\nu\partial^\mu V^\nu -\partial^\mu D^\ast_\nu V^\nu)\bar D^\ast_\mu\nonumber\\
&&\phantom{\mathcal{L}_{D^\ast D^\ast\rho}=}
    +(\partial_\mu D^{\ast\nu}V^\mu
    \bar D^\ast_\nu-D^{\ast\nu} V^\mu\partial_\mu\bar D^\ast_\nu )\big],\\
&&\mathcal{L}_{K^\ast KP}=-ig_{K^\ast KP}
    (\bar K\partial^\mu P-\partial^\mu\bar KP)K_\mu^\ast+\text{H.C.},\\
&&\mathcal{L}_{K^\ast K^\ast P}=
    -g_{K^\ast K^\ast P}\epsilon^{\mu\nu\alpha\beta}
   \partial_\alpha\bar K^\ast_\beta P\partial_\mu K^\ast_\nu ,\\
&&\mathcal{L}_{K^\ast KV}=
    -g_{K^\ast KV}\epsilon^{\eta\tau\rho\sigma}\partial_\rho\bar K^\ast_\sigma \partial_\eta V_\tau K+\text{H.C.},\\
&&\mathcal{L}_{K^\ast K^\ast V}=-ig_{K^\ast K^\ast V}
    \big[
    (\partial^\mu\bar K^\ast_\nu V^\nu-\bar K^\ast_\nu\partial^\mu V^\nu)K^\ast_\mu\nonumber\\
&&\phantom{\mathcal{L}_{K^\ast K^\ast V}=}
    +\bar K^\ast_\mu(\partial^\mu V^\nu K^\ast_\nu
    -V^\nu\partial^\mu K^\ast_\nu )\nonumber\\
&&\phantom{\mathcal{L}_{K^\ast K^\ast V}=}
    +(\bar K^\ast_\nu V^\mu\partial_\mu K^{\ast\nu}
    -\partial_\mu\bar K^\ast_\nu V^\mu K^{\ast\nu})\big],
\end{eqnarray}

where the doublets $D^{(\ast)}$ and $\bar D^{(\ast)}$ are,
\begin{eqnarray}
  D^{(\ast)}=(D^{(\ast)0}, D^{(\ast)+}),\quad \bar D^{(\ast)}=\left(\begin{matrix}
  \bar D^{(\ast)0}\\
  D^{(\ast)-}
  \end{matrix}\right),\label{eq:def-d-matrix}
\end{eqnarray}
The $P$ stands for $\bm\pi$, $\eta$ and $\eta^\prime$, where
\begin{eqnarray}
  \bm\pi=\left(\begin{array}{cc}
  \pi^0 &\sqrt2\pi^+\\
  \sqrt2\pi^-   &-\pi^0\
  \end{array}\right),\label{eq:def-pi}
\end{eqnarray}
and the vector meson $V$ can be $\bm\rho$, $\omega$, where
\begin{eqnarray}
    \bm\pi=\left(\begin{array}{cc}
  \rho^0 &\sqrt2\rho^+\\
  \sqrt2\rho^-   &-\rho^0\
  \end{array}\right).\label{eq:def-rho}
\end{eqnarray}
In our numerical calculations, we simple employ the coupling constants $g_{D^\ast D\pi}=12.2$, which was estimated via the experimental measured decay width of process $D^\ast\to D\pi$\cite{Chen:2014sra}. The $g_{D^\ast D^\ast\pi}=11.9$ was from the Ref.~\cite{Oh:2000qr}. Applying the VMD method to the process $D^\ast\to D\gamma$, one can obtain the $g_{D^\ast D\rho}=2.82$\cite{Oh:2000qr}. The $g_{D^\ast D^\ast \rho}=2.52$ was determined by the same VMD method\cite{Lin:1999ad,Oh:2000qr}. In addition, the coupling constants $g_{K^\ast K\pi}=3.12$ was determined via the experimental measured decay width of the process $K^\ast\to K\pi$\cite{Chen:2011cj}. Moreover, the $g_{K^\ast K\pi}$, $g_{K^\ast K^\ast\pi}$, $g_{K^\ast K\rho}$, and $g_{K^\ast K^\ast\rho}$ can be related via a gauge coupling $g$,
\begin{eqnarray}
&&g_{K^\ast K\pi}=\frac14g,\quad g_{K^\ast K^\ast\pi}=
    \frac{1}{4}\frac{g^2N_c}{16\pi^2F_\pi},\\
&&g_{K^\ast K\rho}=\frac{1}{4}\frac{g^2N_c}{16\pi^2F_\pi},
    \quad g_{K^\ast K^\ast\rho}=\frac{1}{4}g,
\end{eqnarray}
where $N_c=3$ is the number of the quark color, $F_\pi=132$\,MeV is the decay constant of the pion. Other coupling constants concerned to the $\eta^{(\prime)}$ and $\omega$ can be obtained via the $SU(3)$ symmetry.

In terms of the effective lagrangians already given above, we can write out the Feynman amplitudes of the diagrams in Fig.~\ref{fig:tri-x-two-body-decay}. As for the process $X_0\to  D^-K^+$ [Fig.~\ref{fig:tri-x-two-body-decay}-(a)], we have
\begin{eqnarray}
\mathcal{M}_{X_0\to D^-K^+}^P
    &=&\int\frac{d^4q}{(2\pi)^4}\tilde
    \Phi\big[(p_1-w_{12}p)^2\big]
    \big[\frac{1}{\sqrt 2}g_{X_0}\big]\nonumber\\
&&\times
    \big[ig_{D^\ast DP}(iq_\mu)\big]
    \big[-ig_{K^\ast KP}(-iq^\nu-ip_4^\nu)\big]\nonumber\\
&&\times\frac{-g^{\phi\mu}
    +p_1^\phi p_1^\mu/m_1^2}{p_1^2-m_1^2}\frac{-g_{\phi\nu}+p_{2\phi} p_{2\nu}/m_2^2}{p_2^2-m_2^2}\nonumber\\
&&\times
    \frac{1}{q^2-m_q^2}
    \mathcal{F}^2(m_q,\Lambda_1),\\
\mathcal{M}_{X_0\to D^-K^+}^V
    &=&\int\frac{d^4q}{(2\pi)^4}\tilde
    \Phi\big[(p_1-w_{12}p)^2\big]
    \big[\frac{1}{\sqrt 2}g_{X_0}\big]\nonumber\\
&&\times
    \big[g_{D^\ast DV}\epsilon_{\mu\nu\alpha\beta}
    (iq^\mu)(-ip_1^\alpha)\big]
    \big[-g_{K^\ast KV}    \nonumber\\
&&\times
    \epsilon_{\eta\tau\rho\sigma}
    (-iq^\eta)
    (-ip_2^\rho)\big]
    \frac{-g^{\phi\beta}
    +p_1^\phi p_1^\beta/m_1^2}{p_1^2-m_1^2}  \nonumber\\
&&\times
    \frac{-g^{\sigma}_\phi+p_{2\phi} p_2^\sigma/m_2^2}{p_2^2-m_2^2}
    \frac{-g^{\tau\nu}+q^\tau q^\nu/m_q^2}{q^2-m_q^2}\nonumber\\
&&\times\mathcal{F}^2(m_q,\Lambda_1),
\end{eqnarray}
where the $\omega_{12}=m_1/(m_1+m_2)$, the $\mathcal{M}_{X_0\to D^-K^+}^P$ and $\mathcal{M}_{X_0\to D^-K^+}^V$ are the Feynman amplitudes for the transition from $X_0$ to $D^-K^+$ with the exchanging pseudoscalar mesons ($\pi^0,\ \eta,\ \eta^\prime$) and vector mesons ($\rho^0,\ \omega$), respectively.  The mass of the ex-
change meson is $m_q$. Here, a phenomenological form factor was introduced to represent the off-shell effect of the coupling constant, and we selected a dipole form factor\cite{Colangelo:2002mj},
\begin{eqnarray}
\mathcal{F}^2(m_q,\Lambda_1)=
    (\frac{m_q^2-\Lambda_1^2}{q^2-\Lambda_1^2})^2.
\end{eqnarray}

Other two diagrams with the $\bar D^{\ast0}K^{\ast0}$ intermediate states, in which the exchanged states are $\pi^-$ and $\rho^-$, also contribute to the process $X_0\to D^-K^+$, we can obtain the corresponding Feynman amplitude via the isospin symmetry,
\begin{eqnarray}
\mathcal A_{X_0\to D^-K^+}^{\pi^-}&=&-2\mathcal
    M_{X_0\to D^-K^+}^{\pi^0},\\
\mathcal A_{X_0\to D^-K^+}^{\rho^-}&=&-2\mathcal
    M_{X_0\to D^-K^+}^{\rho^0}.
\end{eqnarray}

Similarly, we can write out the Feynman amplitudes for the processes $X_1\to D^{\ast-} K^+$, $X_1\to  D^- K^\ast$, $X_2\to D^-K$, $X_2\to D^{\ast-} K^+$, and $X_2\to D^- K^{\ast+}$. The detailed expressions
were presented in the Appendix.

Now, the total contributions of the processes $X_{J(J=0,1,2)}\to  D^{(\ast)-}K^{(\ast) +}$ were,
\begin{eqnarray}
\mathcal M_{X_{J(J=0,1,2)}\to
     D^{(\ast)-}K^{(\ast) +}}^{\text{tot}}&=&
    \mathcal M^{\pi^0}+\mathcal M^{\eta}
    +\mathcal M^{\eta^\prime}+\mathcal M^{\rho^0}\nonumber\\
&&+\mathcal M^{\omega}\pm(\mathcal A^{\pi^-}+\mathcal A^{\rho^-}),
\end{eqnarray}
where in the right side the lower index $X_{J(J=0,1,2)}\to D^{(\ast)-}K^{(\ast) +}$ of $\mathcal M$ was ignored, the $\pm$ corresponding to $I=0$ and $I=1$ cases, respectively. Finally, we can derive the partial decay widths of the processes $X_{J(J=0,1,2)}\to  D^{(\ast)-}K^{(\ast) +}$,
\begin{eqnarray}
\hspace*{-0.5cm}\Gamma(X_{J(J=0,1,2)}\to D^{(\ast)-}K^{(\ast) +})&=&
    \frac{1}{2J+1}\frac{1}{8\pi}\frac{|\vec p|}{M^2}\nonumber\\
&&\times
    |\bar {\mathcal M}_{X_{J(J=0,1,2)}\to
    D^{(\ast)-}K^{(\ast) +}}^{\text{tot}}|^2,
\end{eqnarray}
where the $J$ and $M$ are the angular momentum and mass of the initial state, respectively, $|\vec p|$ is the three-momentum of the final state in the rest frame of the initial state, the overline represents the sum of the polarization for the  initial and final states.

In terms of the isospin symmetry, the partial decay width of the $\bar D^{(\ast )0}K^{(\ast)0}$ is the same as the $D^{(\ast )-}K^{(\ast)+}$ mode. Therefore,
\begin{eqnarray}
&&\Gamma(X_{J(J=0,1,2)}\to
    \bar D^{(\ast)}K^{(\ast)})\nonumber\\
&&\phantom{\ }=
    \Gamma(X_{J(J=0,1,2)}\to  D^{(\ast)-}K^{(\ast) +})
    +\Gamma(X_{J(J=0,1,2)}\to  \bar D^{(\ast)0}K^{(\ast) 0})\nonumber\\
&&\phantom{\ }=
    2\Gamma(X_{J(J=0,1,2)}\to  D^{(\ast)-}K^{(\ast) +}).
\end{eqnarray}

\section{Numerical Results and discussion}\label{sec:3}

\begin{figure}[hbt!]
\begin{center}
\includegraphics[scale=1.1]{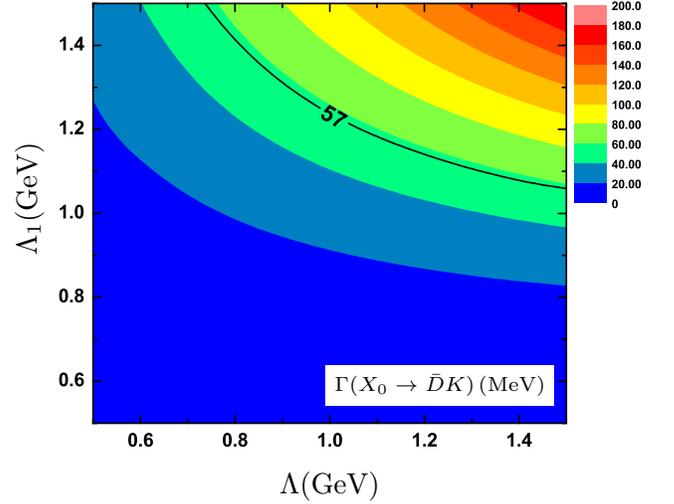}
\end{center}
\caption{The partial decay width of the $X_0\to \bar DK$ with the variation of the parameters $\Lambda$ and $\Lambda_1$, . The solid line corresponding to the center value of  the experimental measured decay width of $X_0$\cite{Aaij:2020ypa}. }\label{fig:num-x0-decay}
\end{figure}
In Fig.~\ref{fig:num-x0-decay}, the numerical results of partial decay width for $X_0\to \bar DK$ process were presented, where the $\Lambda$ and $\Lambda_1$ were the two parameters  in our present approach. Since they cannot be determined by the first principle, the experimental data is usually applied to constrain them. Assuming that the partial decay width of $\D K$ mode was the experimental measured total decay width of $X_0$ resonance, then, the parameters can be constrained via the experimental measured data. On the other hand, we attempted to constrain the parameters within the range $0.5-1.5$\,GeV, while other regions for the cut-off parameters seem unreasonable. The solid line in Fig.~\ref{fig:num-x0-decay} corresponding to the center value of the experimental measured $X_0$ decay width, which is $\Gamma(X_0)=57$\,MeV. Based on this line, a series sets of parameters can be determined, here, we gave serval typical values of the constrained parameter. For $\Lambda=0.8$, 0.9, 1.0, 1.1, 1.2 and 1.3\,GeV, the corresponding $\Lambda_1$ are 1.41, 1.31, 1.24, 1.18, 1.14, 1.11 and 1.08\,GeV, respectively.

\begin{figure}[hbt!]
\begin{center}
\includegraphics[scale=0.85]{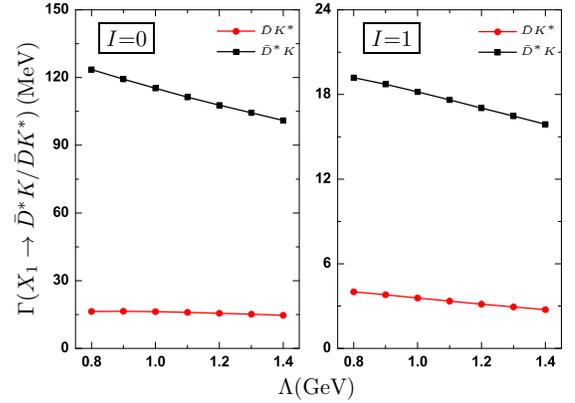}
\end{center}
\caption{The partial decay widths of the $X_1\to \bar D^\ast K$ and $X_1\to \bar D K^\ast$  processes with the constrained parameters. The left column corresponding to the results with $I(X_1)=0$ and the right column was the results with $I(X_1)=1$.  }\label{fig:num-x1-decay}
\end{figure}

The above  typical values of parameters were applied to  predict the decay properties of the $X_1$ and $X_2$, the two spin partners of $X_0(2900)$. Here, both the masses of the $X_1$ and $X_2$ were assigned to be 2866\,MeV, which were predicted in Ref.~\cite{Hu:2020mxp}. In Fig.~\ref{fig:num-x1-decay}, the numerical results of partial decay widths for the $X_1\to\bar D^\ast K$ and $X_1\to\bar D K^\ast$ processes were presented. For the $I({X_1})=0$ case, we found that  $\Gamma(X_1\to \D^\ast K)$  varied from 123.6\,MeV to 101.0\,MeV within the constrained parameters, which weakly depended to the parameters. The partial decay width  for another $\bar DK^{\ast}$ mode varied from 16.4\,MeV to 14.7\,MeV. The numerical results for $I(X_1)=1$ case were much smaller compared to those for $I(X_1)=0$ case, where the partial decay widths for the $\bar D^\ast K$ and $\bar DK^\ast$ were $19.2-15.9$\,MeV and $4.01-2.75$\,MeV, respectively. It could be concluded that  for the both two cases, the $\D^\ast K$ mode was the  dominant decay mode. Besides, we also found that the $X_1$ with $I=0$ was a broad state since the corresponding estimated width was more than 100\,MeV.

\begin{figure}[hbt!]
\begin{center}
\includegraphics[scale=0.85]{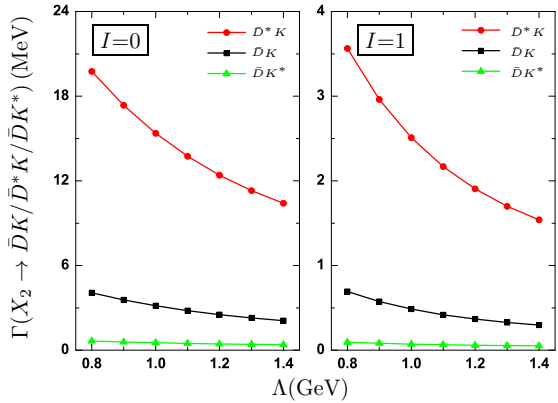}
\end{center}
\caption{The partial decay widths of the transitions from $X_2$ to $\bar DK$, $\D^\ast K$ and $\D K^\ast$ with the constrained  parameters. The left column corresponding to the results with $I(X_2)=0$ and the right column was the results with $I(X_2)=1$. }\label{fig:num-x2-decay}
\end{figure}

As for another $X_2$ state, the numerical results of its partial decay widths were presented in the Fig.~\ref{fig:num-x2-decay}. One can find that the $\bar D^\ast K$ mode was the dominant decay mode both for the $I(X_2)=0$ and $I(X_2)=1$ cases, the corresponding partial decay widths in the constrained parameter range were 19.7$-$10.4\,MeV for the $I(X_2)=0$ case and $3.56-1.54$\,MeV for the $I(X_2)=1$ case. Compared to the $\bar D^\ast K$ mode, the partial decay width for the $\bar D K^\ast$ mode was expected to be, at least, one order of magnitude smaller. In particular, the $\Gamma(X_2\to\bar D K^\ast)$ was $0.649-0.385$\,MeV for the $I(X_2)=0$ case and $0.0921-0.0507$\,MeV for $I(X_2)=1$ case. Besides of the $\bar D^\ast K$ and $\bar DK^\ast$ mode, the $X_2$ can also decay to the $\bar D K$, which was the channel that $X_0$ observed in. The corresponding partial width was $4.06-2.09$\,MeV for the $I(X_2)=0$ case, and $0.692-0.296$\,MeV for the $I(X_2)=1$ case.   Similar to the $X_1$ case, the partial decay widths for the $X_2$ state with $I=0$ was much larger than that that with $I=1$.

\begin{table}[hbt!]
\caption{Predicted partial decay widths for the $X_1$ and $X_2$. The results were based on the typical values of parameters $\Lambda=1.0$\,GeV, $\Lambda_1=1.24$\,GeV.\label{tab:num-decay width}}
\begin{center}
\begin{tabular}{c|cccc}
\hline
\multirow{2}{*}{Partial Decay Width (MeV)}
    &\multicolumn{2}{c}{$X_1$}   &\multicolumn{2}{c}{$X_2$}\\\cline{2-5}
&$I=0$    &$I=1$ &$I=0$    &$I=1$ \\\hline
$\Gamma(\D K)$     &-  &-  &3.15   &0.485\\\hline
$\Gamma(\D^\ast K)$&115   &18.2  &15.4   &2.51\\\hline
$\Gamma(\D K^\ast)$&16.3 &3.58  &0.528 &0.0718\\\hline
\end{tabular}
\end{center}
\end{table}

Based on the above analyses, it was found that the predicted results depended weakly to the parameters. Therefore, in Tab.~\ref{tab:num-decay width}, we also summarized our predictions for the  partial decay widths of $X_1$ and $X_2$  with the typical parameters $\Lambda=1.0$\,GeV and $\Lambda_1=1.24$\,GeV.

\section{Summary}\label{sec:4}

In the present work, we investigated the decay behaviors of $X_0(2900)$ in the $S-$wave $\bar D^\ast K^\ast$ scenario with the isospin $I=0$. With the help of the effective lagrangian approach, the contributions from the triangle diagrams were estimated. Moreover, in order to represent the off-shell effect of the coupling constants,  a phenomenological form factor was considered. The obtained partial decay width for the $X_0\to \bar DK$ process was in agreement with the experimental data with the model parameters $\Lambda$ and $\Lambda_1$ that were selected to be around 1\,GeV.

Within the constrained model parameters, we further calculated the decay behaviors of another two $S-$wave $\bar D^\ast K^\ast$ hadronic molecules $X_1$ and $X_2$, where both the $I=0$ and $I=1$ cases were taken into account. The $X_1$ can decay to $\bar D^\ast K$ and $\bar DK^\ast$, and the $X_2$ can decay to $\bar D K$, $\bar D^\ast K$ and $\bar DK^\ast$. In the constrained parameter ranges, the partial decay widths for the $X_1$ state with $I=0$  were,
\begin{eqnarray}
  \Gamma(X_1\to \bar D^\ast K)&=&124-101\,\text{MeV},\\
  \Gamma(X_1\to \bar D K^\ast)&=&16.4-14.7\,\text{MeV}.
\end{eqnarray}
and for the  $X_2$ state with $I=0$,
\begin{eqnarray}
  \Gamma(X_2\to \bar DK)&=&4.06-2.09\,\text{MeV},\\
  \Gamma(X_2\to \bar D^\ast K)&=&19.8-10.4\,\text{MeV},\\
  \Gamma(X_2\to \bar D K^\ast)&=&0.649-0.385\,\text{MeV}.
\end{eqnarray}
Besides,  we got that the partial decay width for the $I=1$ states were almost one-senventh of that for $I=0$. We concluded that the $X_1$ state with $I=0$ was a broad state with the width more than 100\,MeV, while others were narrow state. Both for the $X_1$ and $X_2$ state, the $\bar D^\ast K$ mode was the dominant decay mode.

Finally, the observation of the $X_0(2900)$ opened a new area for the fully open multi-quark states. The inner structure of the $X_0(2900)$ is  still  controversial. It is valuable to determine the isospin number of $X_0(2900)$ experimentally. Meanwhile, searching for its spin partners and the flavor partners can also help us to understand the nature of $X_0(2900)$. We hoped that more progress can be carried out in the near future.

\section*{Acknowledgement}
This project is supported by the National Key R\&D Program of China under Grant No. 2017YFA0403200, by the National Natural Science Foundation of China under Grant Nos. 11947224, 11975245 and 11775050, by the fund provided to the Sino-German CRC 110 ``Symmetries and the Emergence of
Structure in QCD" project by the NSFC under Grant No.~11621131001, and by the Key Research Program of Frontier Sciences,
CAS, Grant No. Y7292610K1.

\appendix
\section{The amplitudes of the transition from the $X_{J(J=1,2)}$ to $\bar D^{(\ast)}K^{(\ast)}$}

The diagrams contributing to the process $X_{J(J=1,2)}$ to $\bar D^{(\ast)}K^{(\ast)}$ were presented in Fig.~\ref{fig:tri-x-two-body-decay}, we can write out the corresponding Feynman amplitudes.
For the $X_1\to D^{\ast-} K^+$ process,
\begin{eqnarray}
\mathcal{M}_{X_1\to  D^{\ast-} K^+}^P
    &=&\int\frac{d^4q}{(2\pi)^4}\tilde
    \Phi\big[(p_1-w_{12}p)^2\big]\big[\frac{1}{\sqrt 2}g_{X_1}\epsilon_{\kappa\lambda\gamma\theta}(-ip^\kappa)\nonumber\\
&&\times
    \epsilon^\lambda(p_1)\big]
    \big[g_{D^\ast D^\ast P}\epsilon_{\mu\nu\alpha\beta}(ip_3^\mu)(-ip_1^\alpha)
    \epsilon^\nu(p_3)\big]\nonumber\\
&&\times
    \big[-ig_{K^\ast KP}(-iq_\phi-ip_{4\phi})\big]
    \frac{-g^{\gamma\beta}
    +p_1^\gamma p_1^\beta/m_1^2}{p_1^2-m_1^2}\nonumber\\
&&\times
    \frac{-g^{\theta\phi}+p_2^\theta p_2^\phi/m_2^2}{p_2^2-m_2^2}\frac{1}{q^2-m_q^2}\nonumber\\
&&\times\mathcal{F}^2(m_q,\Lambda_1),\\
\mathcal{M}_{X_1\to D^{\ast-} K^+}^V
    &=&\int\frac{d^4q}{(2\pi)^4}\tilde
    \Phi\big[(p_1-w_{12}p)^2\big]\big[\frac{1}{\sqrt 2}g_{X_1}\epsilon_{\kappa\lambda\gamma\theta}(-ip^\kappa)
    \nonumber\\
&&\times
    \epsilon^\lambda(p_1)\big]
    \big\{-ig_{D^\ast D^\ast V}
    \big[(-ip_1^\tau)g^{\eta\rho}-(iq^\tau)g^{\rho\eta} \nonumber\\
&&
    +(iq^\eta)g^{\tau\rho}-(ip_3^\eta)g^{\tau\rho}
    +(ip_3^\rho)g^{\tau\eta}-(ip_1^\rho)g^{\eta\tau}\big]
    \nonumber\\
&&\times
    \epsilon_\tau(p_3)\big\}\big[-g_{K^\ast KV}\epsilon_{\mu\nu\alpha\beta}
    (-iq^\mu)(-ip_2^\alpha)\big]
   \nonumber\\
&&\times
    \frac{-g^{\gamma}_\eta
        +p_1^\gamma p_{1\eta}/m_1^2}{p_1^2-m_1^2}
    \frac{-g^{\theta\beta}
        +p_2^\theta p_2^\beta/m_2^2}{p_2^2-m_2^2}\nonumber\\
&&\times
    \frac{-g^{\nu}_\rho
        +q_\rho q^\nu/m_q^2}{q^2-m_q^2}
        \mathcal{F}^2(m_q,\Lambda_1).
\end{eqnarray}
For the $X_1\to D^-K^{\ast+}$ process,
\begin{eqnarray}
\mathcal{M}_{X_1\to D^-K^{\ast+}}^P
    &=&\int\frac{d^4q}{(2\pi)^4}\tilde
    \Phi\big[(p_1-w_{12}p)^2\big]\big[\frac{1}{\sqrt 2}g_{X_1}\epsilon_{\kappa\lambda\gamma\theta}(-ip^\kappa)
    \nonumber\\
&&\times\epsilon^\lambda(p_1)\big]
    \big[ig_{D^\ast DP}(iq_\mu)\big]
    \big[-g_{K^\ast K^\ast P}\epsilon_{\eta\tau\rho\sigma}\nonumber\\
&&\times(-ip_2^\eta)(ip_4^\rho)
    \epsilon^\sigma(p_4)\big]
    \frac{-g^{\gamma\mu}+p_1^\gamma p_1^\mu/m_1^2}{p_1^2-m_1^2}\nonumber\\
&&\times
    \frac{-g^{\theta\tau}+p_2^\theta p_2^\tau/m_2^2}{p_2^2-m_2^2}\frac{1}{q^2-m_q^2}\nonumber\\
&&\times\mathcal{F}^2(m_q,\Lambda_1),\\
\mathcal{M}_{X_1\to D^-K^{\ast+}}^V
    &=&\int\frac{d^4q}{(2\pi)^4}\tilde
    \Phi\big[(p_1-w_{12}p)^2\big]
    \big[\frac{1}{\sqrt 2}g_{X_1}\epsilon_{\kappa\lambda\gamma\theta}
    (-ip^\kappa)\nonumber\\
&&\times
    \epsilon^\lambda(p_1)\big]
    \big[g_{D^\ast D V}\epsilon_{\mu\nu\alpha\beta}(iq^\mu)(-ip_1^\alpha)
    \big]
    \big\{-ig_{K^\ast K^\ast V}\nonumber\\
&&\times
    \big[(-iq^\rho)g^{\tau\eta}-(-iq^\eta)g^{\rho\tau}
 +(-ip_2^\tau)g^{\rho\eta}\nonumber\\
&&-(-ip_2^\rho)g^{\tau\eta}
    +(ip_4^\eta)g^{\tau\rho}-(ip_4^\tau)g^{\eta\rho}\big]
    \epsilon_\rho(p_4)
    \big\}\nonumber\\
&&\times\frac{-g^{\gamma\beta}
        +p_1^\gamma p_1^\beta/m_1^2}{p_1^2-m_1^2}
    \frac{-g^{\theta}_\eta
        +p_2^\theta p_{2\eta}/m_2^2}{p_2^2-m_2^2}
        \nonumber\\
&&\times
    \frac{-g_\tau^{\nu}
        +q_\tau q^\nu/m_q^2}{q^2-m_q^2}
        \mathcal{F}^2(m_q,\Lambda_1).
\end{eqnarray}
For the $X_2\to D^-K^+$ process,
\begin{eqnarray}
\mathcal{M}_{X_2\to D^-K^+}^P
    &=&\int\frac{d^4q}{(2\pi)^4}\tilde
    \Phi\big[(p_1-w_{12}p)^2\big]
    \big[\frac{1}{\sqrt 2}g_{X_2}
    \epsilon_{\kappa\lambda}(p)\big]
    \nonumber\\
&&\times
    \big[ig_{D^\ast D P}(-iq_\mu)
    \big]
    \big[-ig_{K^\ast KP}
    (-iq_\nu\nonumber\\
&&-ip_{4\nu})\big]
    \frac{-g^{\kappa\mu}
        +p_1^\kappa p_1^\mu/m_1^2}{p_1^2-m_1^2}
    \frac{-g^{\lambda\nu}
        +p_2^\lambda p_2^\nu/m_2^2}{p_2^2-m_2^2}\nonumber\\
&&\times\frac{1}{q^2-m_q^2}
        \mathcal{F}^2(m_q,\Lambda_1),\\
\mathcal{M}_{X_2\to D^-K^+}^V
    &=&\int\frac{d^4q}{(2\pi)^4}\tilde
    \Phi\big[(p_1-w_{12}p)^2\big]
    \big[\frac{1}{\sqrt 2}g_{X_2}
    \epsilon_{\kappa\lambda}(p)\big]\nonumber\\
&&\times
    \big[g_{D^\ast D V}\epsilon_{\mu\nu\alpha\beta}
    (-iq^\mu)(-ip_1^\alpha)\big]
    \big[-g_{K^\ast KV}
    \epsilon_{\eta\tau\rho\sigma}\nonumber\\
&&\times(-iq^\eta)(-ip_2^\rho)\big]
    \frac{-g^{\kappa\beta}
        +p_1^\kappa p_1^\beta/m_1^2}{p_1^2-m_1^2}\nonumber\\
&&\times
    \frac{-g^{\lambda\sigma}
        +p_2^\lambda p_2^\sigma/m_2^2}{p_2^2-m_2^2}
    \frac{-g^{\tau\nu}
        +q^\tau q^\nu/m_q^2}{q^2-m_q^2}\nonumber\\
&&\times
        \mathcal{F}^2(m_q,\Lambda_1).
\end{eqnarray}
For the $X_2\to D^{\ast-}K^+$ process,
\begin{eqnarray}
\mathcal{M}_{X_2\to D^{\ast-} K^+}^P
    &=&\int\frac{d^4q}{(2\pi)^4}\tilde
    \Phi\big[(p_1-w_{12}p)^2\big]
    \big[\frac{1}{\sqrt 2}g_{X_2}
    \epsilon_{\kappa\lambda}(p)\big]\nonumber\\
&&\times
    \big[g_{D^\ast D^\ast P}\epsilon_{\mu\nu\alpha\beta}
    (ip_3^\mu)(-ip_1^\alpha)\epsilon^\nu(p_3)
    \big]\nonumber\\
&&\times
    \big[-ig_{K^\ast KP}(-iq_\sigma-ip_{4\sigma})\big]\nonumber\\
&&\times
    \frac{-g^{\kappa\beta}
        +p_1^\kappa p_1^\beta/m_1^2}{p_1^2-m_1^2}
    \frac{-g^{\lambda\sigma}
        +p_2^\lambda p_2^\sigma/m_2^2}{p_2^2-m_2^2}
       \nonumber\\
&&\times \frac{1}{q^2-m_q^2}\mathcal{F}^2(m_q,\Lambda_1),\\
\mathcal{M}_{X_2\to D^{\ast-} K^+}^V
    &=&\int\frac{d^4q}{(2\pi)^4}\tilde
    \Phi\big[(p_1-w_{12}p)^2\big]
    \big[\frac{1}{\sqrt 2}g_{X_2}
    \epsilon_{\kappa\lambda}(p)\big]\nonumber\\
&&\times
    \big\{-ig_{D^\ast D^\ast V}
    \big[(-ip_1^\tau)g^{\eta\rho}-(iq^\tau)g^{\rho\eta}\nonumber\\
&&+(iq^\eta)g^{\tau\rho}-(ip_3^\eta)g^{\tau\rho}
    +(ip_3^\rho)g^{\tau\eta}\nonumber\\
&&-(-ip_1^\rho)g^{\eta\tau}\big]
    \epsilon_\tau(p_3)\big\}
    \big[-g_{K^\ast KV}
    \epsilon_{\mu\nu\alpha\beta}(-iq^\mu)\nonumber\\
&&\times
    (-ip_2^\alpha)\big]
    \frac{-g^{\kappa}_{\eta}
        +p_1^\kappa p_{1\eta}/m_1^2}{p_1^2-m_1^2}\frac{-g^{\lambda\beta}
        +p_2^\lambda p_2^\beta/m_2^2}{p_2^2-m_2^2}\nonumber\\
&&\times
    \frac{-g^{\nu}_{\rho}
        +q^\nu q_\rho/m_q^2}{q^2-m_q^2}
        \mathcal{F}^2(m_q,\Lambda_1).
\end{eqnarray}
For the process $X_2\to D^- K^{\ast+}$
\begin{eqnarray}
\mathcal{M}_{X_2\to D^- K^{\ast+}}^P
    &=&\int\frac{d^4q}{(2\pi)^4}\tilde
    \Phi\big[(p_1-w_{12}p)^2\big]\big[\frac{1}{\sqrt 2}g_{X_2}\epsilon_{\kappa\lambda}(p)\big]\nonumber\\
&&\times
    \big[ig_{D^\ast DP}(iq_\mu)\big]
    \big[-g_{K^\ast K^\ast P}\epsilon_{\eta\tau\rho\sigma}\nonumber\\
&&\times(-ip_2^\eta)(ip_4^\rho)
    \epsilon^\sigma(p_4)\big]
    \frac{-g^{\kappa\mu}+p_1^\kappa p_1^\mu/m_1^2}{p_1^2-m_1^2}\nonumber\\
&&\times
    \frac{-g^{\lambda\tau}+p_2^\lambda p_2^\tau/m_2^2}{p_2^2-m_2^2}\frac{1}{q^2-m_q^2}\nonumber\\
&&\times\mathcal{F}^2(m_q,\Lambda_1),\\
\mathcal{M}_{X_2\to D^-K^{\ast+}}^V
    &=&\int\frac{d^4q}{(2\pi)^4}\tilde
    \Phi\big[(p_1-w_{12}p)^2\big]
    \big[\frac{1}{\sqrt 2}g_{X_2}
    \epsilon_{\kappa\lambda}(p)\big]\nonumber\\
&&\times
    \big[g_{D^\ast D V}\epsilon_{\mu\nu\alpha\beta}(iq^\mu)(-ip_1^\alpha)
    \big]
    \big\{-ig_{K^\ast K^\ast V}\nonumber\\
&&\times
    \big[(-iq^\rho)g^{\tau\eta}-(-iq^\eta)g^{\rho\tau}
    +(-ip_2^\tau)g^{\rho\eta}\nonumber\\
&&-(-ip_2^\rho)g^{\tau\eta}
    +(ip_4^\eta)g^{\tau\rho}-(ip_4^\tau)g^{\eta\rho}\big]
    \epsilon_\rho(p_4)\big\}
    \nonumber\\
&&\times
    \frac{-g^{\kappa\beta}
        +p_1^\kappa p_1^\beta/m_1^2}{p_1^2-m_1^2}
        \frac{-g^{\lambda}_{\eta}
        +p_2^\lambda p_{2\eta}/m_2^2}{p_2^2-m_2^2}\nonumber\\
&&\times
    \frac{-g_{\tau}^{\nu}
        +q_\tau q^\nu/m_q^2}{q^2-m_q^2}
        \mathcal{F}^2(m_q,\Lambda_1).
\end{eqnarray}

\end{document}